\documentclass{vgtc}                          

\graphicspath{{figures/}{pictures/}{images/}{./}} 

\usepackage{times}                     

\usepackage{tabu}                      
\usepackage{booktabs}                  
\usepackage{lipsum}                    
\usepackage{mwe}                       
\usepackage{amsmath}
\usepackage{amssymb}
\usepackage[hang,flushmargin]{footmisc} 
\usepackage[normalem]{ulem}
\usepackage[scaled=0.85]{FiraMono}

\usepackage{tikz}
\usepackage{fancyvrb}
\usepackage[T1]{fontenc}
\usepackage{enumerate}
\usepackage[pagebackref,bookmarks,hyperfootnotes=false]{hyperref}

\usepackage{mathptmx}                  
\usepackage[sort,numbers]{natbib}      

\usepackage[hyphenation]{impnattypo}
\newcommand{\customsubsubsection}[1]{%
    \smallskip\noindent%
       \fontfamily{cmss}{\textbf{#1}}%
       \normalfont
}

\definecolor{myblue}{HTML}{9588F7}

\onlineid{0}

\vgtccategory{Research}

\vgtcinsertpkg

\newcommand{\ccdf}[1]{mCCDF#1}

\def\trackchanges{0}

\newcommand{\new}[1]{%
\if\trackchanges1%
\textcolor{red}{#1}%
\else#1\fi}
\newcommand{\stout}[1]{\if\trackchanges1 \textcolor{red}{\sout{#1}} \fi}

\newcommand{\supplement}{\href{https://osf.io/cdgby/}{\texttt{supplement}}}

\title{Adapting CCDF Plots for Visualizing Ordinal Regression Results}

\author{Abhraneel Sarma\thanks{e-mail: asarma@tugraz.at}\\ %
        \scriptsize Graz University of Technology}

\abstract{
    Cumulative-link ordinal regression models are an alternative approach for analysing ordinal data such as Likert items, which are widely used in Visualization (and other related fields like \textsc{hci}, psychology etc.). There are many researchers who are strong proponents of this approach, as it makes less stringent assumptions about the data, compared to the more commonly used linear model or \textsc{anova}. Yet, ordinal regression models have seen limited adoption. I posit that \new{one possible reason for this might be} due to the difficulty in visually representing the results from such models, and in communicating the key takeaways in an intuitive manner. I propose the use of (modified) Complementary Cumulative Distribution Function (\ccdf) plots to visualize the results of ordinal regression models, and demonstrate how the same takeaways that researchers present from analyses which treat ordinal data as metric can be easily communicated using \ccdf{s}.
} 

\keywords{Complementary Cumulative Distribution Functions, Likert scale, Ordinal Regression Model}

\begin{document}


\firstsection{Introduction}

\maketitle

An ordinal variable refers to variables where the categories have a natural order, but are not metric~\cite{stevens_theory_1946} (i.e., the differences between successive levels may not be treated as equal intervals). The use of ordinal measures (e.g., surveys which elicit participants' responses on a Likert scale) are commonly used by researchers in visualization, as well as in other fields such as \textsc{hci} or psychology. These measures provide a valuable way of measuring metrics such as trust in a visualization, confidence in using a new interface, preference or attitudes etc., which we wouldn't be able to capture otherwise.

Despite their widespread use, researchers tend to use a wide variety of statistical approaches for analysing Likert data, some of which makes strong and arguably unfounded assumptions about the data. For instance, the \textsc{anova} test, frequently used for analysing Likert responses, assumes that the data is metric~\cite{liddell_analyzing_2018}. Others use non-parametric approaches which, while not making assumptions on the data distribution, reduces the data to relative ordered ranks~\cite{victorsyiem_better_2026} which may also be problematic. Instead, Syiem and Velloso~\cite{victorsyiem_better_2026}, echoing prior calls from other fields~\cite{liddell_analyzing_2018}, have argued for the adoption of ordinal regression models instead, as these models make less stringent assumptions regarding the data, and more directly model the data generating process. 

Like Syiem and Velloso~\cite{victorsyiem_better_2026}, I also believe that ordinal cumulative-link models are the more appropriate approach for analysing Likert data---ordinal models make much more reasonable assumptions regarding the data-generating process and generally tends to fit the data better.%
\footnote{However, how often ordinal models lead to different conclusions in practice is unclear, and I elaborate more on this in Section~\ref{sec:is-ordinal-better}}
However, I posit that the a key reason for the lack of adoption of ordinal models is 
{\stout{perhaps \textit{not} due to researchers not knowing how to implement such models, as there have been numerous blogposts~\cite[e.g.,][]{medina_ordinal_2021, kurz_ordinal_2026, kurz_causal_2023}, books~\cite[e.g.,][]{mcelreath_statistical_2020} and scientific articles~\cite[e.g.,][]{burkner_ordinal_2019, gambarota_ordinal_2024, agresti_tutorial_1989, victorsyiem_better_2026} describing both frequentist and Bayesian approaches; instead,}}
\new{
the fact that \textbf{we currently do not have good approaches for visualising the results of ordinal models} which are effective at communicating the takeaways that analysts wish to communicate regarding their analysis.} As prior work has used various approaches for communicating the results of Likert scale responses, I first conduct a preliminary review of reporting strategies used for ordinal data, based on a snowball sample of prior work; I summarise the key message that researchers are trying to communicate and identify key differences between when the data is treated as metric or ordinal.%
\footnote{I do not look at non-parametric responses because I think that approach represents the worst of both worlds---i.e., makes different unfounded assumptions (i.e., by converting the data into ranks, it flattens the difference between a ordered dataset of \{5, 2, 3\} and \{7, 2, 3\} as both get ranked as \{3, 1, 2\}~\cite{victorsyiem_better_2026}, and remains confusing to interpret.}

I find that an important benefit of a metric-based analysis of ordinal responses is the ability to report results and describe differences between conditions on the response scale (e.g., participants in condition [$X$], on average, reported 2 points higher trust on a 7-point Likert scale, compared to participants in condition [$Y$]); these differences can also be very intuitively visualized (see \autoref{sec:reporting-strategies}). Analogous statements are very difficult (although not impossible) to make if the data were analyzed using an ordinal model instead. Based on these findings, I propose the use of modified Complementary Cumulative Distribution Function (\ccdf) plots to visualize the results of ordinal regression models. Even though \ccdf{s} are less straightforward for interpreting the results (compared to the interval plots that researchers use when treating the results as metric), they do not violate the ordinal assumption by treating the data as metric. More importantly, I demonstrate how \ccdf{s} allow an analyst to communicate the analogous information using ordinal models that metric-based analyses of ordinal data affords, which can be further reinforced through annotations (\autoref{sec:ccdf}).

\section{Preliminaries}
To make the discussion throughout this paper more concrete, I implement and use the results of an ordinal regression model. I introduce the dataset and the the model below (see also \autoref{appendix:model-details})

\vspace{-2pt}
\subsection{The Dataset}
To demonstrate the \ccdf{} visualization, we first need an ordinal model. I use a dataset on moral intuition~\cite{cushman_role_2006, mcelreath_statistical_2020} which contains the results of an experiment where participants are presented with different scenarios involving the \textit{trolley problem}. The scenarios manipulate three principles---\textit{action}, \textit{intention} and \textit{contact}.
In this experiment, participants are presented with a number of scenarios where one or more of these principles apply; thus whether a principle applies in a particular scenario is the experimental variable. Participants' responses are recorded on a 7-point scale where higher values indicate greater moral permissibility. For a more detailed discussion on the experiment, I would point readers to~\cite{mcelreath_statistical_2020, cushman_role_2006}.

\begin{figure*}[t!]
    \includegraphics[width=\textwidth]{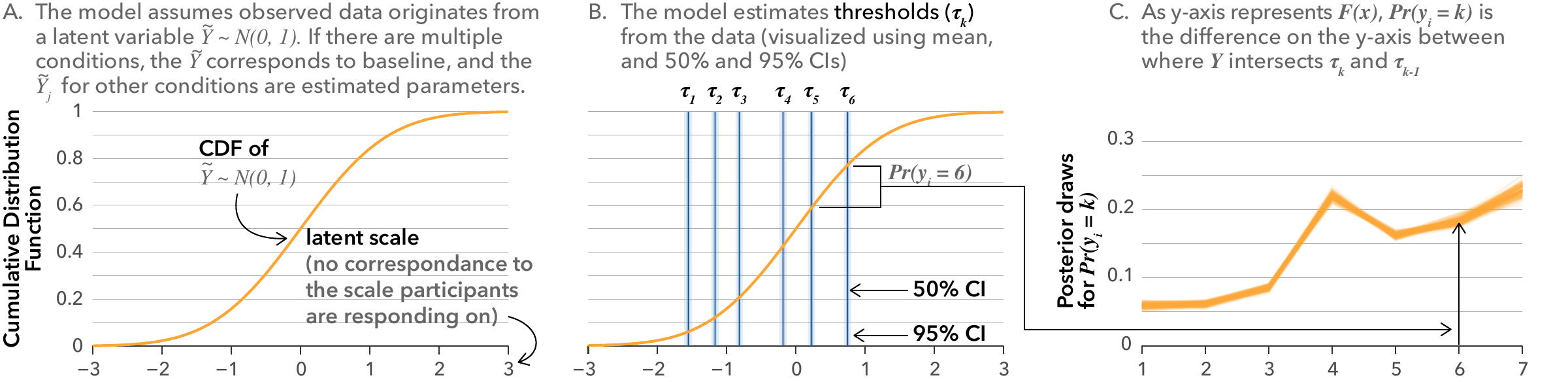}
    \vspace{-20pt}
    \caption{Illustration of how a cumulative probit-link model estimates thresholds of a latent variable (B), which can then be used to estimate (posterior) probabilities for the prevalance of each rating (C).%
    }
    \label{fig:ordinal-regression-process}
    \vspace{-10pt}
\end{figure*}

\vspace{-2pt}
\subsection{The Ordinal Regression Model}
I fit a Bayesian%
\footnote{\new{The complete analysis can be found in \supplement{} $\blacktriangleright$ RScript $\blacktriangleright$ \texttt{01-bayesian.Rmd}. I also show how to create the corresponding visualizations for a frequentist analysis in \supplement{} $\blacktriangleright$ RScript $\blacktriangleright$ \texttt{02-frequentist.Rmd}}}
cumulative-link model using the probit link function on the \textit{moral intuition} dataset using \texttt{action}, \texttt{intention} and \texttt{contact} as predictors. The cumulative model assumes that an observed ordinal variable, $Y$, originates from a categorization of a latent continuous variable, $\tilde{Y}$~\cite{burkner_ordinal_2019, victorsyiem_better_2026}, which follows the unit normal distribution (\autoref{fig:ordinal-regression-process}A). To achieve this categorization, the model then assumes that there are $k$ thresholds $\tau_k$ which categorize the latent variable $\tilde{Y}$ into $k + 1$ \textit{ordered} categories (\autoref{fig:ordinal-regression-process}B). The thresholds $\tau_k$, are on the scale of the latent variable, and are estimated from the data:

\vspace{-6pt}
\[ \Phi^{-1}(\mathbb{P}(y_i \leq k)) = \tau_k \]
\vspace{-12pt}

\noindent where $\Phi$ is the probit function%
\footnote{\new{
A logit function can also be used as a link function in cumulative-link models. The role of the link function is to provide a one-to-one mapping between a continuous variable in $\mathbb{R}$ to $[0, 1]$. In theory, there are many such functions $f: \mathbb{R} \rightarrow [0, 1]$ that can be used as link functions. The choice of which link function to use is essentially arbitrary~\cite{yang_subjective_2023}, and typically chosen based on convention and ease of interpretability. Here, I use the probit function due to mathematical convenience.
}}
corresponding to the unit normal distribution. We can estimate the probability of $Y$ being equal to a category $k$ as $\mathbb{P}(Y = k) = \Phi^{-1}(\tau_k) - \Phi^{-1}(\tau_{k-1})$. This is extended to a regression model by assuming that the different conditions will have different underlying latent distributions $\tilde{Y_j}$ (see \autoref{fig:ordinal-regression-process}).

\section{Visualizing Ordinal Data}
\label{sec:reporting-strategies}

To get a better understanding of what researchers want to communicate from their analysis of ordinal data, I conducted an informal survey of prior work in visualization.
I began with a few initial publications that I was familiar with which contained analyses of ordinal data~\cite[e.g.,][]{sarma_evaluating_2023, yang_dice_2024, yang_swaying_2024, nadib_guardrail_2026, lisnic_visualization_2025} and \new{conducted a search in the \textsc{ieee} digital library using either the keyword \textit{likert} or \textit{ordinal regression} for papers published between the years 2025--2026 at \textsc{ieee vis}, \new{the premier publication venue in the field of visualization} (similar to the methodology used by South et al~\cite{south_effective_2022}). This search returned 60 results. However, as this yielded only two examples of studies which used ordinal regression models, I expanded the years to 2020--2026 only for the keyword \textit{ordinal regression}. This resulted in a sample of 72 papers, of which 55 used a visualization to communicate the results of a Likert data analysis. I supplement the findings from my survey with the results of South et al.'s~\cite{south_effective_2022} systematic review of the use of Likert scale in Visualization papers. The qualitative analysis can be found in \supplement{} $\blacktriangleright$ survey.}
\stout{However, this returned only one example of a study which used ordinal regression models, suggesting that treating ordinal data as metric is likely the much more common approach; thus, I conducted a subsequent search using the keyword \textit{ordinal regression} for papers published at \textsc{ieee vis} between the years 2020 -- 2026.}%
I chose this informal approach, as the objective of this survey was to get a sense of the diversity in analysis and reporting strategies that researchers have used \new{and how researchers communicate the key takeaways from their Likert data}, and not to quantify how frequently each modelling approach has been adopted. 

\begin{figure*}[t]
    \includegraphics[width=\textwidth]{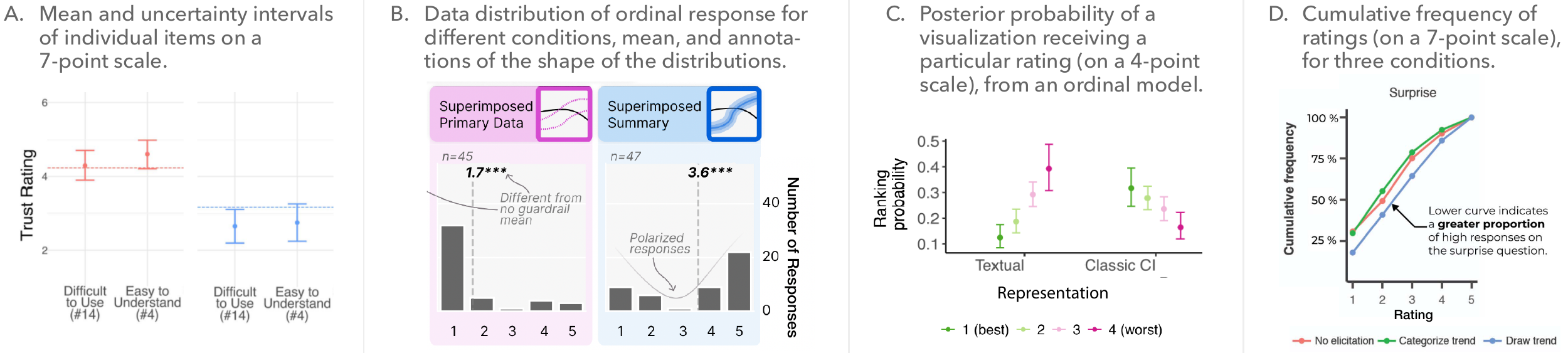}
    \vspace{-22pt}
    \caption{
    Prototypical examples of visualizations of ordinal data analyses from the survey. (A) shows mean and error bars representing standard errors (but other visualizations show 95\% confidence or credible intervals) of individual items~\cite{wang_you_2026}. Similar visualizations were used to represent estimates aggregated across multiple items~\cite[e.g.,][]{cabouat_previs_2025, yang_dice_2024, yang_swaying_2024}.
    (B) shows the complete data distribution with annotations of the mean and shape of the distribution~\cite{lisnic_visualization_2025}.
    (C) shows the posterior probability of a condition receiving a particular rating on a 4-point scale~\cite{helske_can_2021}.
    (D) Empirical cumulative distribution function of ratings (on a 7-point scale) for three treatment conditions~\cite{rogha_impact_2024}. Instead of using empirical frequencies, a similar visualization could also be created using posterior draws from an ordinal regression model.
    }
    \label{fig:reporting-ordinal-data}
    \vspace{-12pt}
\end{figure*}

\subsection{When the Data is Modeled as Metric}
Several of the papers in my survey (25/55) reported means and an uncertainty estimate (most commonly 95\% confidence or credible intervals, but also in some cases standard deviations; see~\autoref{fig:reporting-ordinal-data}A) for each item~\cite[e.g.,][]{song_visualizing_2026, nadib_guardrail_2026, wang_you_2026}. Some researchers report the means and 95\% intervals aggregated across multiple items~\cite[e.g.,][]{cabouat_previs_2025, yang_dice_2024}, by taking the average for each participant, which is how Likert scales were original intended to be used~\cite{likert_technique_1932, carifio_ten_2007}. Other approaches that I observed were reporting the data distribution directly (\autoref{fig:reporting-ordinal-data}B) and comparing the means and the shapes of distribution.~\cite[e.g.,][]{lisnic_visualization_2025}, or using boxplots and densities~\cite[e.g.,][]{nakano_avatars_2025}. \new{In South et al's review~\cite{south_effective_2022}, 30/65 papers with a visualization communicate some form of mean and uncertainty estimate}.

Researchers primarily used ordinal data to make comparisons between conditions%
\stout{i.e., which condition is rated higher.}
by communicating mean point estimates and differences in means on the response scale. Even though a (hypothetical) statement such as ``\textit{participants rated visualization [$X$] to be more trustworthy (M = 4.1, SD = 1.05) compared to visualization [$Y$] (M = 3.6, SD = 1.4)}'' may be technically flawed\footnote{Researchers also use statistical significance from \textsc{anova}s to support their claims about which conditions are more highly rated, which is arguably more problematic---statistical significance tests should be conducted in conjunction with a power analysis, lest a researcher risks running overpowered studies, and many of the studies do. However, it is unclear whether the assumptions of power analyses for commonly used statistical tests hold for such skewed data distributions which are of concern here (e.g., \textsc{anova} assumes that the data is normally distributed). Additionally, the ritualistic use of statistical significance tests as rituals~\cite{gigerenzer_mindless_2004, gigerenzer_statistical_2018} is a broader issue which is out of scope of this current work.}
when applied to data which is not on an interval scale as it entails arithmetic operations~\cite{victorsyiem_better_2026}, it provides a sense of the data distribution and the average response in a particular condition are (at least approximately), and crucially how large the differences between conditions are \textit{on the response scale}. For instance, Nadib et al.~\cite{nadib_guardrail_2026} reporting higher trust in one of their conditions contextualize this by adding ``the effect, albeit statistically significant, is not large: trust increases by roughly half of a Likert point [...].''

\subsection{When the Data is Modeled as Ordinal}
I observed four distinct approaches for communicating the results of an ordinal model: (i) visualizing the probability of each ordinal category for each condition~\cite{sarma_evaluating_2023, saske_multidimensional_2026, helske_can_2021}; (ii) visualizing the empirical cumulative distribution function (\textsc{cdf}) plots~\cite{rogha_impact_2024}; (iii) showing the data distribution and calculating the odds ratio between conditions~\cite{bradley_magnitude_2025}; and (iv) estimating mean and 95\% credible intervals from posterior draws (for a single item or averaged across multiple items)~\cite{yang_swaying_2024, dragicevic_increasing_2019}. The difference in how difficult it is to comprehend the results of ordinal models compared to metric models is stark---the first two approaches (\autoref{fig:reporting-ordinal-data}C-D), while more appropriate for the analysis, can make it significantly more difficult for a viewer%
\stout{to quickly and intuitively make sense of what is going on} 
\new{to determine how much higher/better one condition is rated on average, compared to other conditions, in terms of points on the Likert scale (a straightforward task using the mean and interval plots).}

This might be perhaps why Yang et al.~\cite{yang_swaying_2024} and Dragicevic et al.~\cite{dragicevic_increasing_2019} adopted the approach of first modelling the data as ordinal, and then treating it as metric by estimating the mean of posterior predicted probability\footnote{In Dragicevic et al.~\cite{dragicevic_increasing_2019}, the posterior probability for each rating is multiplied by the ``magnitude'' of the rating, $\sum_k \mathbb{P}(y_i = k) \cdot k$, to obtain the posterior mean and credible interval} across the ratings---it allowed them to communicate results on the response scale. While the approach by Rogha et al.~\cite{rogha_impact_2024} (see \autoref{fig:reporting-ordinal-data}D) visualizes the empirical cumulative distribution function (i.e., the probability that a participant rated at least [$k$] on the 7-point Likert scale) for each condition on the response scale, it remains challenging and counter-intuitive to interpret\new{---to determine which condition is rated higher, the viewer has to identify the curve which is \textit{visually lower}, which runs counter to the typical convention in visualization  of encoding higher as better}. 
This unintuitiveness of the ordinal model has also led to an arguably idiosyncratic decision for reporting results in my own prior work~\cite{sarma_evaluating_2023}---I reported the probability of participants rating \textit{3 or higher} on a 5-point Likert scale to compare across conditions, with the choice of 3 or higher being arbitrary (although hopefully justifiable). More conventionally, Bradley et al.'s~\cite{bradley_magnitude_2025} decision to report the odds ratio is similar to recommendations made by Syiem and Velloso~\cite{victorsyiem_better_2026}; they even augment this by converting the odd's ratio to a Cohen's d value. However, despite all this, the odds ratio between two variables on a latent (inverse probit) scale is significantly less intuitive than information communicated on the response scale.

\new{Researchers also often communicated their key takeaways based on the proportion of responses that fell into, or above, a specific category, without modelling the data as either metric or ordinal, and using visualizations of the distribution of Likert responses such as histograms, stacked bar charts or heatmaps (26/55 papers in my survey and 27/65 papers in~\cite{south_effective_2022}). This takeaway, which is analogous to estimating $Pr(y_i \geq k)$, can be easily obtained using an ordinal model; however, this value can be quite difficult to estimate using either of the visualizations used in prior work (\autoref{fig:reporting-ordinal-data}C-D).}

\section{Towards a Better Visualization of Ordinal Regression Models}
\label{sec:ccdf}

\subsection{Design Properties}
Based on the discussion above, I outline a set of design properties for visualizing the results of ordinal regression models which will make it intuitive for a viewer to easily extract the key pieces of information they care about:

\customsubsubsection{[D1]} The visualization should communicate the central tendency on the response scale (e.g., what is the model estimated response for the median participant?

\customsubsubsection{[D2]} The visualization should communicate uncertainty in the central tendency on the response scale (e.g., what is the uncertainty in the median estimate?

\customsubsubsection{[D3]} The visualization should easily allow the reader to identify which condition is rated ``better.'' Consistent with most convention, I argue that the visual representation should encode conditions which are ``better'' higher (unlike the \textsc{cdf} plot in \autoref{fig:reporting-ordinal-data}D).

\customsubsubsection{[D4]} The visualization should allow comparison between conditions on the response scale (e.g., how many points higher does the median participant in one condition rate than the median participant in another condition?).


\noindent \new{I propose the use of a \textit{modified} Complementary Cumulative Distribution Function, which fulfils these design properties, as a better visualization of the results of ordinal models}

\begin{figure}[b!]
    \vspace{-12pt}
    \includegraphics[width=\columnwidth]{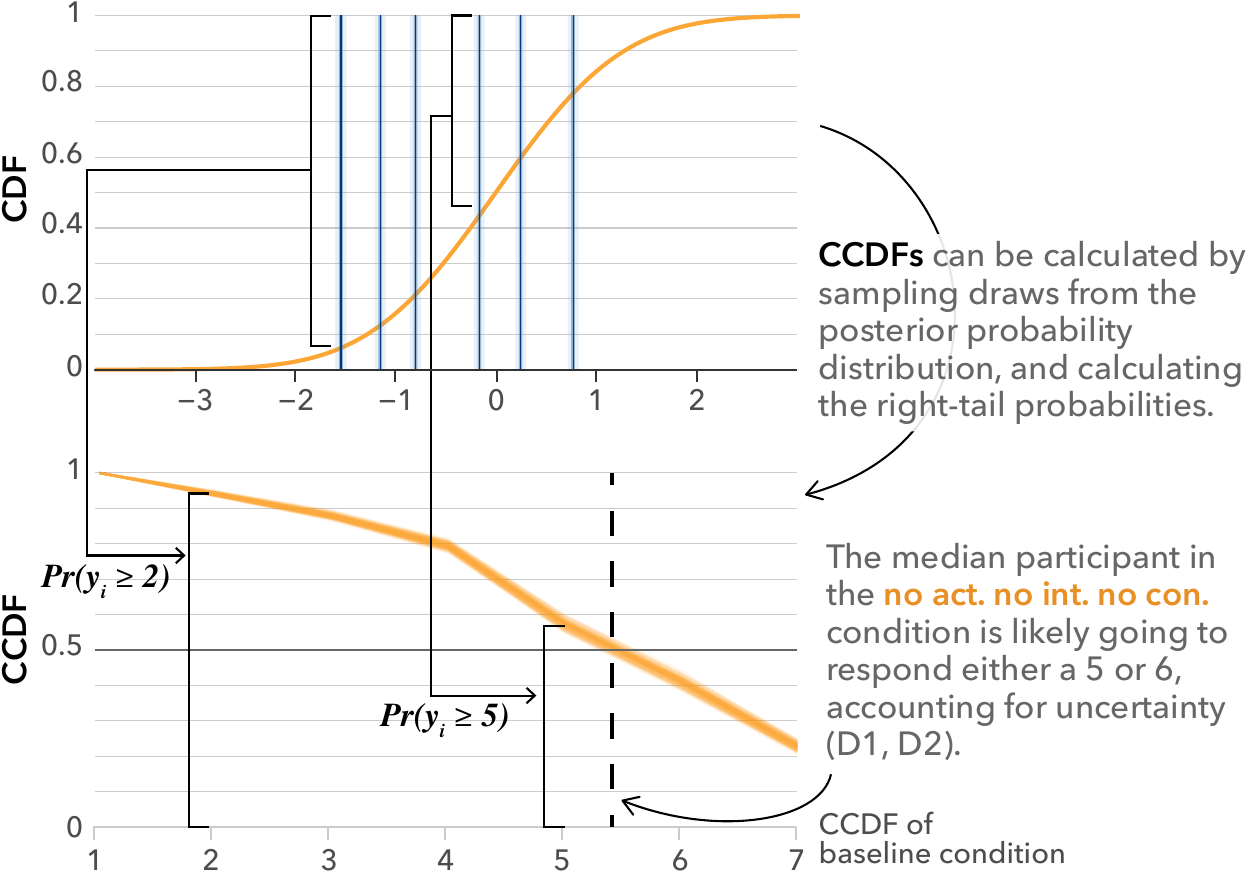}
    \vspace{-16pt}
    \caption{Calculation of a \ccdf{} plot from the latent variable and estimated thresholds.}
    \label{fig:ccdf-calc}
\end{figure}

\subsection{The modified Complementary Cumulative Distribution Function (mCCDF) Plot}

The Complementary Cumulative Distribution Function (CCDF) can be calculated from a cumulative probit-link regression model using a linear transformation: $\mathrm{CCDF} = \Phi^{-1}(y_i > k) = 1 - \Phi^{-1}(y_i \leq k)$. \new{Thus, the \textsc{ccdf} tells us the probability of an item being rated greater than category $k$. However, for Likert data, we care about the probability of an item being rated at category $k$ or higher, which can be obtained using the following modified CCDF}:

\vspace{-6pt}
\[ \mathrm{mCCDF} = \Phi^{-1}(y_i {\color{myblue}{\geq}} k) = 1 - \Phi^{-1}(y_i < k) \]
\vspace{-12pt}

\noindent While seemingly trivial to estimate, the \ccdf{} has properties which make it more suitable to communicate the results of ordinal regression models, especially for single items. The rating corresponding to the intersection of the \ccdf{} draws with $y = 0.5$ represents the median rating for a particular condition. In~\autoref{fig:ccdf-calc}, the median rating is likely going to be a 5 or 6, even after taking uncertainty into account (D1, D2). The uncertainty in the model estimates is relatively low for this model and dataset due to the large number of responses collected.

\begin{figure}[t!]
    \vspace{3pt}
    \includegraphics[width=\columnwidth]{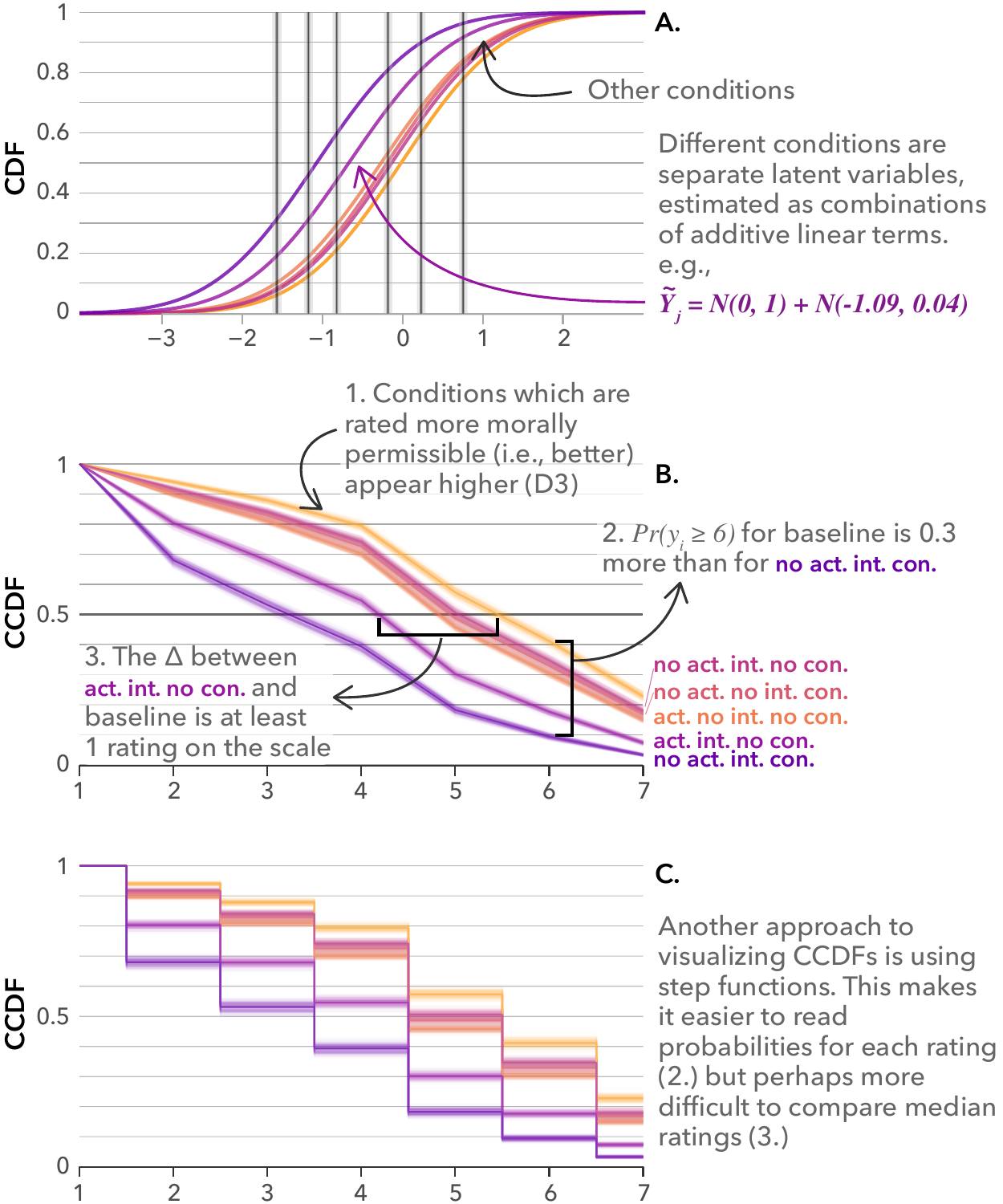}
    \vspace{-18pt}
    \caption{Calculation of a \ccdf{} plot for an experiment with multiple conditions (i.e., ordinal models with multiple predictors).}
    \label{fig:ccdf-conditions}
    \vspace{-13pt}
\end{figure}

In addition, the \ccdf{} more directly encodes one of the most relevant pieces of information for ordinal data---the probability that participants rated $k$ or higher ($\mathbb{P}(y_i \geq k)$), unlike the conventional \textsc{cdf} plot which encodes $\mathbb{P}(y_i \leq k)$. This makes it more intuitive to perform two types of comparisons: (i) how much more likely are participants in one condition to rate $k$ or higher relative to another \new{(D3), as shown in \autoref{fig:ccdf-conditions}B2} (e.g., the probability of the \textit{action, intention, no contact} condition being rated 6 or higher is 0.3 less than the baseline condition); and (ii) how much better one condition is relative to another \new{(D4)}, based on the median participant (e.g., in \autoref{fig:ccdf-conditions}B(3), the median participant in the \textit{action, intention, no contact} condition is going to rate at least a four, and the median participant in the baseline condition is going to rate at least a 5). Note that while these comparisons can be performed by a viewer using \textsc{cdf} plots as well, they are far less intuitive (for an approximation of the challenge, compare the magnitude and direction of the difference between conditions in \autoref{fig:ccdf-conditions}A).

An alternative approach for representing \ccdf{s} is visualizing them as step functions (a piecewise constant function). Representing \ccdf{s} as step functions has some trade-offs---reading the mean probabilities for each rating (and the associated uncertainty) for a particular condition becomes easier; however, determining the median rating for a condition, and thus comparing the median rating between conditions becomes more difficult. As such, the designer should choose which communication goal they wish to prioritise.
\footnote{I personally tend to prefer the line plot over the step function plot.}

\subsection{Scales with Multiple Correlated Items}

In some of the studies in the survey~\cite{cabouat_previs_2025, yang_swaying_2024, yang_dice_2024} responses from multiple, potentially correlated, items were aggregated and reported using a single point estimate and uncertainty interval.%
\stout{
The analogous approach which treats the data as ordinal would be to take the average, across the multiple scales, of the (posterior) probabilities for each rating
}
\new{
An analyst could instead fit a single hierarchical model to the data, with the correlated items modelled as random effects with different latent means and variances (referred to as \textit{difficulty} and \textit{discrimination} parameters in item-response theory). As scales with potentially correlated items are often measuring the same underlying construct, partial pooling of information in a hierarchical model allows parameters to be more efficiently estimated. We can then create mCCDF plots for each item individually, an ``average'' item (i.e., when the latent mean is zero), or marginalized over all of the items. I demonstrate how such an analysis can be implemented, and how the corresponding mCCDF plots can be created for each item in \supplement{} $\blacktriangleright$ RScript $\blacktriangleright$ \texttt{03-unequal-variance-ccdf.Rmd}.
}

\new{
An additional benefit of adopting such models is that they can be useful for evaluating the items themselves, by assessing the degree of similarity in responses to the different items. Analogous approaches are used in item-response theory to evaluate questions, based on the coefficients for the difficulty and discrimination parameters and using item characteristic curves (ICC), to create validated set of test questions (see Section 5-7 and Figure 11 in~\cite{ge_calvi_2023}). For ordinal models, mCCDFs can complement the information presented using ICCs (which show the modeled probability of responding at or above each category as a function of the latent trait) by visualizing how similar participants' actual responses to each item actually are, in the response scale.
}


\subsection{Should We All Be Using Ordinal Models?}
\label{sec:is-ordinal-better}
While I personally prefer to use ordinal models for Likert data, I do not hold strong beliefs that this is objectively the best way of analysing ordinal data, as I am yet to find compelling evidence to suggest that the conclusions would be different regardless of the choice of models. Liddell and Kruschke~\cite{liddell_analyzing_2018} try to demonstrate that ordinal regression is better, and that treating ordinal data as metric can lead to inflated false positive and false negatives, using simulated data. However, for their simulation, they generate ordinal data by discretizing a latent variable; thus, what they actually demonstrated was that the model which is consistent with their assumed data-generating process was better than a model which made obviously incorrect assumptions in the \textit{small world} of their simulation.\footnote{The true data generating process is almost always unknown, and the models that we use to analyze data are \textit{all} small world approximations~\cite{savage_foundations_1972}, and thus rely on assumptions. The uncertainty regarding which model is correct has been described as ontological uncertainty~\cite{spiegelhalter_risk_2017, sarma_milliways_2024, sarma_more_2025}.} In contrast, Dragicevic et al.~\cite{dragicevic_increasing_2019} compare nine different data analysis approaches for the same dataset and found little difference in terms of takeaways between ordinal regression and metric-based approaches. 
\stout{
Thus, the objective of this work is not necessarily to urge researchers to adopt ordinal regression models, but rather demonstrate approaches for visualizing results of ordinal regression \textit{if} researchers choose to adopt such models.
}
\new{My preference for using ordinal models is one based on statistical philosophy---I consider approaches which model the data directly, with as few assumptions as possible, to be better. The value of an ordinal model lies in the fact that it makes fewer assumptions regarding the data generating processes of Likert responses compared to metric-based approaches, which is why I would still recommend the use of ordinal regression model. However, the primary goal of this paper is not necessarily to urge researchers to adopt ordinal regression models but rather to demonstrate approaches for visualizing results of ordinal regression \textit{if} researchers choose to adopt such models.}

\section{Conclusion}
When analysing ordinal data, researchers appear to prefer to treat the data as metric and communicate the results---using point estimates, uncertainty, and differences between conditions---on the response scale. Complementary Cumulative Distribution Function (\ccdf) plots offer an approach for communicating the results of cumulative-link ordinal regression models on the response scale, allowing the reader to extract analogous pieces of information.



\acknowledgments{
I would like to thank Matthew Kay for detailed feedback on the visualization and the paper draft.
}

\bibliographystyle{abbrv-doi}

\bibliography{template}

@article{agresti_tutorial_1989,
  title = {Tutorial on Modeling Ordered Categorical Response Data},
  author = {Agresti, Alan},
  year = 1989,
  journal = {Psychological Bulletin},
  volume = {105},
  number = {2},
  pages = {290--301},
  publisher = {American Psychological Association},
  address = {US},
  issn = {1939-1455},
  doi = {10.1037/0033-2909.105.2.290}
}

@article{bradley_magnitude_2025,
  title = {Magnitude {{Judgements}} Are {{Influenced}} by the {{Relative Positions}} of {{Data Points Within Axis Limits}}},
  author = {Bradley, Duncan and Strain, Gabriel and Jay, Caroline and Stewart, Andrew J.},
  year = 2025,
  month = feb,
  journal = {IEEE Transactions on Visualization and Computer Graphics},
  volume = {31},
  number = {2},
  pages = {1414--1421},
  issn = {1941-0506},
  doi = {10.1109/TVCG.2024.3364069},
  urldate = {2026-04-29}
}

@article{burkner_ordinal_2019,
  title = {Ordinal {{Regression Models}} in {{Psychology}}: {{A Tutorial}}},
  shorttitle = {Ordinal {{Regression Models}} in {{Psychology}}},
  author = {B{\"u}rkner, Paul-Christian and Vuorre, Matti},
  year = 2019,
  month = mar,
  journal = {Advances in Methods and Practices in Psychological Science},
  volume = {2},
  number = {1},
  pages = {77--101},
  publisher = {SAGE Publications Inc},
  issn = {2515-2459},
  doi = {10.1177/2515245918823199},
  urldate = {2026-04-27},
  langid = {english}
}

@article{cabouat_previs_2025,
  title = {{{PREVis}}: {{Perceived Readability Evaluation}} for {{Visualizations}}},
  shorttitle = {{{PREVis}}},
  author = {Cabouat, Anne-Flore and He, Tingying and Isenberg, Petra and Isenberg, Tobias},
  year = 2025,
  month = jan,
  journal = {IEEE Transactions on Visualization and Computer Graphics},
  volume = {31},
  number = {1},
  pages = {1083--1093},
  issn = {1941-0506},
  doi = {10.1109/TVCG.2024.3456318},
  urldate = {2026-04-29}
}

@article{carifio_ten_2007,
  title = {Ten {{Common Misunderstandings}}, {{Misconceptions}}, {{Persistent Myths}} and {{Urban Legends}} about {{Likert Scales}} and {{Likert Response Formats}} and Their {{Antidotes}}},
  author = {Carifio, James and Perla, Rocco J.},
  year = 2007,
  month = mar,
  journal = {Journal of Social Sciences},
  volume = {3},
  number = {3},
  pages = {106--116},
  issn = {15493652},
  doi = {10.3844/jssp.2007.106.116},
  urldate = {2026-04-29},
  langid = {english}
}

@article{cushman_role_2006,
  title = {The {{Role}} of {{Conscious Reasoning}} and {{Intuition}} in {{Moral Judgment}}: {{Testing Three Principles}} of {{Harm}}},
  shorttitle = {The {{Role}} of {{Conscious Reasoning}} and {{Intuition}} in {{Moral Judgment}}},
  author = {Cushman, Fiery and Young, Liane and Hauser, Marc},
  year = 2006,
  month = dec,
  journal = {Psychological Science},
  volume = {17},
  number = {12},
  pages = {1082--1089},
  issn = {0956-7976, 1467-9280},
  doi = {10.1111/j.1467-9280.2006.01834.x},
  urldate = {2026-05-01},
  copyright = {https://journals.sagepub.com/page/policies/text-and-data-mining-license},
  langid = {english}
}

@inproceedings{dragicevic_increasing_2019,
  title = {Increasing the {{Transparency}} of {{Research Papers}} with {{Explorable Multiverse Analyses}}},
  booktitle = {Proceedings of the 2019 {{CHI Conference}} on {{Human Factors}} in {{Computing Systems}}},
  author = {Dragicevic, Pierre and Jansen, Yvonne and Sarma, Abhraneel and Kay, Matthew and Chevalier, Fanny},
  year = 2019,
  month = may,
  pages = {1--15},
  publisher = {ACM},
  address = {Glasgow Scotland Uk},
  doi = {10.1145/3290605.3300295},
  urldate = {2026-04-27},
  isbn = {978-1-4503-5970-2},
  langid = {english}
}

@article{gambarota_ordinal_2024,
  title = {Ordinal Regression Models Made Easy: {{A}} Tutorial on Parameter Interpretation, Data Simulation and Power Analysis},
  shorttitle = {Ordinal Regression Models Made Easy},
  author = {Gambarota, Filippo and Alto{\`e}, Gianmarco},
  year = 2024,
  month = dec,
  journal = {International Journal of Psychology},
  volume = {59},
  number = {6},
  pages = {1263--1292},
  publisher = {John Wiley \& Sons, Ltd},
  issn = {1464-066X},
  doi = {10.1002/ijop.13243},
  urldate = {2026-04-27},
  langid = {english}
}

@inproceedings{ge_calvi_2023,
author = {Ge, Lily W. and Cui, Yuan and Kay, Matthew},
title = {CALVI: Critical Thinking Assessment for Literacy in Visualizations},
year = {2023},
isbn = {9781450394215},
publisher = {Association for Computing Machinery},
address = {New York, NY, USA},
url = {https://doi.org/10.1145/3544548.3581406},
doi = {10.1145/3544548.3581406},
booktitle = {Proceedings of the 2023 CHI Conference on Human Factors in Computing Systems},
articleno = {815},
numpages = {18},
location = {Hamburg, Germany},
series = {CHI '23}
}

@article{gigerenzer_mindless_2004,
  title = {Mindless Statistics},
  author = {Gigerenzer, Gerd},
  year = 2004,
  month = nov,
  journal = {The Journal of Socio-Economics},
  volume = {33},
  number = {5},
  pages = {587--606},
  issn = {10535357},
  doi = {10.1016/j.socec.2004.09.033},
  urldate = {2026-04-29},
  langid = {english}
}

@article{gigerenzer_statistical_2018,
  title = {Statistical {{Rituals}}: {{The Replication Delusion}} and {{How We Got There}}},
  shorttitle = {Statistical {{Rituals}}},
  author = {Gigerenzer, Gerd},
  year = 2018,
  month = jun,
  journal = {Advances in Methods and Practices in Psychological Science},
  volume = {1},
  number = {2},
  pages = {198--218},
  publisher = {SAGE Publications Inc},
  issn = {2515-2459},
  doi = {10.1177/2515245918771329},
  urldate = {2026-04-29},
  langid = {english}
}

@article{helske_can_2021,
  title = {Can {{Visualization Alleviate Dichotomous Thinking}}? {{Effects}} of {{Visual Representations}} on the {{Cliff Effect}}},
  shorttitle = {Can {{Visualization Alleviate Dichotomous Thinking}}?},
  author = {Helske, Jouni and Helske, Satu and Cooper, Matthew and Ynnerman, Anders and Besan{\c c}on, Lonni},
  year = 2021,
  month = aug,
  journal = {IEEE Transactions on Visualization and Computer Graphics},
  volume = {27},
  number = {8},
  pages = {3397--3409},
  issn = {1941-0506},
  doi = {10.1109/TVCG.2021.3073466},
  urldate = {2026-04-29}
}

@misc{kurz_causal_2023,
  title = {Causal Inference with Ordinal Regression},
  author = {Kurz, A. Solomon},
  year = 2023,
  month = may,
  urldate = {2026-04-27},
  howpublished = {https://solomonkurz.netlify.app/blog/2023-05-21-causal-inference-with-ordinal-regression/},
  langid = {english}
}

@misc{kurz_ordinal_2026,
  title = {11 {{Monsters}} and {{Mixtures}} -- {{Statistical}} Rethinking with Brms, Ggplot2, and the Tidyverse},
  author = {Kurz, A. Solomon},
  year = 2026,
  month = jan,
  journal = {*Statistical rethinking* with brms, ggplot2, and the tidyverse},
  urldate = {2026-04-27},
  howpublished = {https://solomon.quarto.pub/sr/11.html},
  langid = {english}
}

@article{liddell_analyzing_2018,
  title = {Analyzing Ordinal Data with Metric Models: {{What}} Could Possibly Go Wrong?},
  shorttitle = {Analyzing Ordinal Data with Metric Models},
  author = {Liddell, Torrin M. and Kruschke, John K.},
  year = 2018,
  month = nov,
  journal = {Journal of Experimental Social Psychology},
  volume = {79},
  pages = {328--348},
  issn = {00221031},
  doi = {10.1016/j.jesp.2018.08.009},
  urldate = {2026-04-27},
  langid = {english}
}

@article{likert_technique_1932,
  title = {A Technique for the Measurement of Attitudes},
  author = {Likert, R.},
  year = 1932,
  journal = {Archives of Psychology},
  volume = {22  140},
  pages = {55--55}
}

@inproceedings{lisnic_visualization_2025,
  title = {Visualization {{Guardrails}}: {{Designing Interventions Against Cherry-Picking}} in {{Interactive Data Explorers}}},
  shorttitle = {Visualization {{Guardrails}}},
  booktitle = {Proceedings of the 2025 {{CHI Conference}} on {{Human Factors}} in {{Computing Systems}}},
  author = {Lisnic, Maxim and Cutler, Zach and Kogan, Marina and Lex, Alexander},
  year = 2025,
  month = apr,
  pages = {1--19},
  publisher = {ACM},
  address = {Yokohama Japan},
  doi = {10.1145/3706598.3713385},
  urldate = {2026-04-29},
  isbn = {979-8-4007-1394-1},
  langid = {english}
}

@book{mcelreath_statistical_2020,
  title = {Statistical {{Rethinking}}: {{A Bayesian Course}} with {{Examples}} in {{R}} and {{STAN}}},
  shorttitle = {Statistical {{Rethinking}}},
  author = {McElreath, Richard},
  year = 2020,
  month = mar,
  edition = {2},
  publisher = {{Chapman and Hall/CRC}},
  address = {New York},
  doi = {10.1201/9780429029608},
  isbn = {978-0-429-02960-8}
}

@misc{medina_ordinal_2021,
  title = {Ordinal Visualization},
  author = {Medina, Octavio},
  year = 2021,
  month = jul,
  urldate = {2026-04-27},
  howpublished = {https://octavio.me/posts/ordinal-viz/},
  langid = {english}
}

@article{nadib_guardrail_2026,
  title = {Guardrail {{Selection}} in {{Line Charts}} to {{Contextualize Persuasive Visualizations}}},
  author = {Nadib, K A and Kogan, M and Lex, A and Lisnic, M},
  year = {2026},
  journal = {Computer Graphics Forum},
  langid = {english}
}

@article{nakano_avatars_2025,
  title = {Avatars, {{Should We Look}} at {{Them Directly}} or {{Through}} a {{Mirror}}?: {{Effects}} of {{Avatar Display Method}} on {{Sense}} of {{Embodiment}} and {{Gaze}}},
  shorttitle = {Avatars, {{Should We Look}} at {{Them Directly}} or {{Through}} a {{Mirror}}?},
  author = {Nakano, Kizashi and Narumi, Takuji},
  year = 2025,
  month = may,
  journal = {IEEE Transactions on Visualization and Computer Graphics},
  volume = {31},
  number = {5},
  pages = {2912--2922},
  issn = {1941-0506},
  doi = {10.1109/TVCG.2025.3549545},
  urldate = {2026-04-29}
}

@article{rogha_impact_2024,
  title = {The {{Impact}} of {{Elicitation}} and {{Contrasting Narratives}} on {{Engagement}}, {{Recall}} and {{Attitude Change With News Articles Containing Data Visualization}}},
  author = {Rogha, Milad and Sah, Subham and Karduni, Alireza and Markant, Douglas and Dou, Wenwen},
  year = 2024,
  month = jul,
  journal = {IEEE Transactions on Visualization and Computer Graphics},
  volume = {30},
  number = {7},
  pages = {4375--4389},
  issn = {1941-0506},
  doi = {10.1109/TVCG.2024.3355884},
  urldate = {2026-04-29}
}

@article{sarma_evaluating_2023,
  title = {Evaluating the {{Use}} of {{Uncertainty Visualisations}} for {{Imputations}} of {{Data Missing At Random}} in {{Scatterplots}}},
  author = {Sarma, Abhraneel and Guo, Shunan and Hoffswell, Jane and Rossi, Ryan and Du, Fan and Koh, Eunyee and Kay, Matthew},
  year = 2023,
  month = jan,
  journal = {IEEE Transactions on Visualization and Computer Graphics},
  volume = {29},
  number = {1},
  pages = {602--612},
  issn = {1941-0506},
  doi = {10.1109/TVCG.2022.3209348},
  urldate = {2026-04-27}
}

@inproceedings{sarma_milliways_2024,
author = {Sarma, Abhraneel and Hwang, Kyle and Hullman, Jessica and Kay, Matthew},
title = {Milliways: Taming Multiverses through Principled Evaluation of Data Analysis Paths},
year = {2024},
isbn = {9798400703300},
publisher = {Association for Computing Machinery},
address = {New York, NY, USA},
url = {https://doi.org/10.1145/3613904.3642375},
doi = {10.1145/3613904.3642375},
booktitle = {Proceedings of the 2024 CHI Conference on Human Factors in Computing Systems},
articleno = {607},
numpages = {15},
location = {Honolulu, HI, USA},
series = {CHI '24}
}

@inproceedings{sarma_more_2025,
  title = {More {{Forecasts}}, {{More}} ({{Decision}}) {{Problems}}: {{How Uncertainty Representations}} for {{Multiple Forecasts Impact Decision Making}}},
  shorttitle = {More {{Forecasts}}, {{More}} ({{Decision}}) {{Problems}}},
  booktitle = {Proceedings of the 2025 {{CHI Conference}} on {{Human Factors}} in {{Computing Systems}}},
  author = {Sarma, Abhraneel and Hedayati, Maryam and Kay, Matthew},
  year = 2025,
  month = apr,
  series = {{{CHI}} '25},
  pages = {1--14},
  publisher = {Association for Computing Machinery},
  address = {New York, NY, USA},
  doi = {10.1145/3706598.3713725},
  urldate = {2026-05-01},
  isbn = {979-8-4007-1394-1}
}

@article{saske_multidimensional_2026,
  title = {A {{Multidimensional Assessment Method}} for {{Visualization Understanding}} ({{MdamV}})},
  author = {Saske, Antonia and Koesten, Laura and M{\"o}ller, Torsten and Staudner, Judith and Kritzinger, Sylvia},
  year = 2026,
  month = mar,
  journal = {IEEE Transactions on Visualization and Computer Graphics},
  volume = {32},
  number = {3},
  pages = {2695--2708},
  issn = {1941-0506},
  doi = {10.1109/TVCG.2026.3653265},
  urldate = {2026-04-29}
}

@book{savage_foundations_1972,
  title = {The Foundations of Statistics},
  author = {Savage, Leonard J.},
  year = 1972,
  edition = {2d rev. ed},
  publisher = {Dover Publications},
  address = {New York},
  isbn = {978-0-486-62349-8},
  langid = {english},
  lccn = {HA29 .S28 1972}
}

@article{song_visualizing_2026,
  title = {Visualizing {{Trust}}: {{How Chart Embellishments Influence Perceptions}} of {{Credibility}}},
  shorttitle = {Visualizing {{Trust}}},
  author = {Song, Hayeong and Cho, Aeree and Bearfield, Cindy Xiong and Stasko, John},
  year = 2026,
  month = jan,
  journal = {IEEE Transactions on Visualization and Computer Graphics},
  volume = {32},
  number = {1},
  pages = {1306--1316},
  issn = {1941-0506},
  doi = {10.1109/TVCG.2025.3634785},
  urldate = {2026-04-29}
}

@article{south_effective_2022,
author = {South, Laura and Saffo, David and Vitek, Olga and Dunne, Cody and Borkin, Michelle A.},
title = {Effective Use of Likert Scales in Visualization Evaluations: A Systematic Review},
journal = {Computer Graphics Forum},
volume = {41},
number = {3},
pages = {43-55},
doi = {https://doi.org/10.1111/cgf.14521},
url = {https://onlinelibrary.wiley.com/doi/abs/10.1111/cgf.14521},
eprint = {https://onlinelibrary.wiley.com/doi/pdf/10.1111/cgf.14521},
year = {2022}
}

@article{spiegelhalter_risk_2017,
  title = {Risk and {{Uncertainty Communication}}},
  author = {Spiegelhalter, David},
  year = 2017,
  month = mar,
  journal = {Annual Review of Statistics and Its Application},
  volume = {4},
  number = {Volume 4, 2017},
  pages = {31--60},
  publisher = {Annual Reviews},
  issn = {2326-8298, 2326-831X},
  doi = {10.1146/annurev-statistics-010814-020148},
  urldate = {2025-05-22},
  langid = {english}
}

@article{stevens_theory_1946,
  title = {On the {{Theory}} of {{Scales}} of {{Measurement}}},
  author = {Stevens, S. S.},
  year = 1946,
  month = jun,
  journal = {Science},
  volume = {103},
  number = {2684},
  pages = {677--680},
  publisher = {American Association for the Advancement of Science},
  doi = {10.1126/science.103.2684.677},
  urldate = {2026-04-27}
}

@inproceedings{victorsyiem_better_2026,
  title = {Better {{Assumptions}}, {{Stronger Conclusions}}: {{The Case}} for {{Ordinal Regression}} in {{HCI}}},
  shorttitle = {Better {{Assumptions}}, {{Stronger Conclusions}}},
  booktitle = {Proceedings of the 2026 {{CHI Conference}} on {{Human Factors}} in {{Computing Systems}}},
  author = {Victor Syiem, Brandon and Velloso, Eduardo},
  year = 2026,
  month = apr,
  pages = {1--21},
  publisher = {ACM},
  address = {Barcelona Spain},
  doi = {10.1145/3772318.3790821},
  urldate = {2026-04-19},
  isbn = {979-8-4007-2278-3},
  langid = {english}
}

@article{wang_you_2026,
  title = {Do {{You}} ``{{Trust}}'' {{This Visualization}}? {{An Inventory}} to {{Measure Trust}} in {{Visualizations}}},
  shorttitle = {Do {{You}} ``{{Trust}}'' {{This Visualization}}?},
  author = {Wang, Huichen Will and Lin, Kylie and Cohen, Andrew and Kennedy, Ryan and Zwald, Zachary and Nobre, Carolina and Bearfield, Cindy Xiong},
  year = 2026,
  month = mar,
  journal = {IEEE Transactions on Visualization and Computer Graphics},
  volume = {32},
  number = {3},
  pages = {2515--2528},
  issn = {1941-0506},
  doi = {10.1109/TVCG.2025.3646847},
  urldate = {2026-04-29}
}

@inproceedings{yang_subjective_2023,
  title = {Subjective {{Probability Correction}} for {{Uncertainty Representations}}},
  booktitle = {Proceedings of the 2023 {{CHI Conference}} on {{Human Factors}} in {{Computing Systems}}},
  author = {Yang, Fumeng and Hedayati, Maryam and Kay, Matthew},
  year = 2023,
  month = apr,
  series = {{{CHI}} '23},
  pages = {1--17},
  publisher = {Association for Computing Machinery},
  address = {New York, NY, USA},
  doi = {10.1145/3544548.3580998},
  urldate = {2026-06-19},
  isbn = {978-1-4503-9421-5}
}

@inproceedings{yang_dice_2024,
  title = {In {{Dice We Trust}}: {{Uncertainty Displays}} for {{Maintaining Trust}} in {{Election Forecasts Over Time}}},
  shorttitle = {In {{Dice We Trust}}},
  booktitle = {Proceedings of the {{CHI Conference}} on {{Human Factors}} in {{Computing Systems}}},
  author = {Yang, Fumeng and Mortenson, Chloe Rose and Nisbet, Erik and Diakopoulos, Nicholas and Kay, Matthew},
  year = 2024,
  month = may,
  pages = {1--24},
  publisher = {ACM},
  address = {Honolulu HI USA},
  doi = {10.1145/3613904.3642371},
  urldate = {2025-04-08},
  isbn = {979-8-4007-0330-0},
  langid = {english}
}

@article{yang_swaying_2024,
  title = {Swaying the {{Public}}? {{Impacts}} of {{Election Forecast Visualizations}} on {{Emotion}}, {{Trust}}, and {{Intention}} in the 2022 {{U}}.{{S}}. {{Midterms}}},
  shorttitle = {Swaying the {{Public}}?},
  author = {Yang, Fumeng and Cai, Mandi and Mortenson, Chloe and Fakhari, Hoda and Lokmanoglu, Ayse D. and Hullman, Jessica and Franconeri, Steven and Diakopoulos, Nicholas and Nisbet, Erik C. and Kay, Matthew},
  year = 2024,
  month = jan,
  journal = {IEEE Transactions on Visualization and Computer Graphics},
  volume = {30},
  number = {1},
  pages = {23--33},
  issn = {1941-0506},
  doi = {10.1109/TVCG.2023.3327356},
  urldate = {2025-04-08}
}

@article{bates_fitting_2015,
title = {Fitting {Linear} {Mixed}-{Effects} {Models} {Using} lme4},
volume = {67},
copyright = {Copyright (c) 2015 Douglas Bates, Martin Mächler, Ben Bolker, Steve Walker},
issn = {1548-7660},
url = {https://doi.org/10.18637/jss.v067.i01},
doi = {10.18637/jss.v067.i01},
language = {en},
urldate = {2023-09-04},
journal = {Journal of Statistical Software},
author = {Bates, Douglas and Mächler, Martin and Bolker, Ben and Walker, Steve},
month = oct,
year = {2015},
pages = {1--48}
}

@article{pinheiro2017package,
  title={nlme : Linear and nonlinear mixed effects models. R package version 3.1-103},
  author={Pinheiro, Jos{\'e} and Bates, Douglas and DebRoy, Saikat and Sarkar, Deepayan and Heisterkamp, Siem and Van Willigen, Bert and Maintainer, R},
  journal={http://cran.r-project.org/web/packages/nlme/index.html},
  year={2017},
  URL={https://cir.nii.ac.jp/crid/1570854174288831360}
}

@article{wilkinson1973symbolic,
  title={Symbolic description of factorial models for analysis of variance},
  author={Wilkinson, GN and Rogers, CE},
  journal={Journal of the Royal Statistical Society: Series C (Applied Statistics)},
  volume={22},
  number={3},
  pages={392--399},
  year={1973},
  doi = {https://doi.org/10.2307/2346786},
  publisher={Wiley Online Library}
}

@misc{RLang_2024,
  title = {R: {{A Language}} and {{Environment}} for {{Statistical Computing}}},
  author = {R Core Team},
  year = 2024,
  address = {Vienna, Austria},
  howpublished = {R Foundation for Statistical Computing}
}

@misc{cmdstanr,
  title = {Cmdstanr: {{R Interface}} to '{{CmdStan}}'},
  author = {Gabry, Jonah and {\v C}e{\v s}novar, Rok and Johnson, Andrew and Bronder, Steve},
  year = 2025
}

\appendix

\section{Supplementary Materials}
\label{appendix:supp}
All supplemental materials are available on OSF at \url{https://osf.io/cdgby/}, released under a CC BY 4.0 license. In particular, they include (1) Excel files containing the data for the literature survey, (2) full R scripts for recreating the analysis and figures in the paper, (3) figure images, and (4) a full version of this paper with all appendices.

\section{Details on Dataset and Ordinal Model Used}
\label{appendix:model-details}

\subsection{Dataset}
The dataset used in the analysis described in this paper was collected by Cushman et al.~\cite{cushman_role_2006} in a series of experiments to collect empirical evidence on how people judge the ``moral goodness or badness'' of actions~\cite{mcelreath_statistical_2020}. The experiments present participants with a scenario---variations of the trolley problem---and asks participants to rate the action (or inaction) on a scale from 1 to 7. The following is the traditional version of the trolley problem~\cite{mcelreath_statistical_2020}:

\begin{quote}
``Standing by the railroad tracks, Dennis sees an empty, out-of-control trolley about to hit five people. Next to Dennis is a lever that can be pulled, sending the trolley down a side track and away from the five people. But pulling the lever will also lower the railing on a footbridge spanning the side track, causing one person to fall off the footbridge and onto the side track, where he will be hit by the trolley. If Dennis pulls the lever the trolley will switch tracks and not hit the five people, and the one person to fall and be hit by the trolley. If Dennis does not pull the lever the trolley will continue down the tracks and hit five people, and the one person will remain safe above the side track.''
\end{quote}

The experiments by Cushman et al.~\cite{cushman_role_2006} investigate the role of three principles from moral psychology and philosophy on moral judgements. The principles are:

\vspace{-4pt}
\begin{enumerate}
    \itemsep0pt
    \item \textbf{The action principle}: Harm caused by action is morally worse than equivalent harm caused by omission.
    \item \textbf{The contact principle}: Using physical contact to cause harm to a victim is morally worse than causing equivalent harm to a victim without using physical contact.
    \item \textbf{The intention principle}: Harm intended as the means to a goal is morally worse than equivalent harm foreseen as the side effect of a goal.
\end{enumerate}
\vspace{-4pt}

The above scenario applies the \textit{action} principle as the actor has to perform an action to create the outcome (as opposed to being passive and letting events unfold). However, the other two principles do not apply as there is no contact (i.e., no \textit{contact}), and the harm caused to the person on the footbridge is accidental and not intentional (i.e., no \textit{intention}). The three principles can be varied to create different scenarios. In the study by Cushman et al.~\cite{cushman_role_2006}, participants ``received 32 moral scenarios separated into two blocks of 16. Each block included 15 test scenarios and 1 control scenario.'' The dataset contains responses from 331 participants, and includes details on participants' age, gender, and level of education. For more details on the dataset, please refer to~\cite{mcelreath_statistical_2020, cushman_role_2006}.

\subsection{Analysis}
The \ccdf{} visualizations described in this paper are based on the results of the analysis outlined by McElreath~\cite{mcelreath_statistical_2020}, which applies a Bayesian ordinal cumulative-logit model. However, I instead use the probit link function (as opposed to the logit link function used by McElreath~\cite{mcelreath_statistical_2020}), due to mathematical convenience. The model can be described using the Wilkinson-Rogers-Pinheiro-Bates syntax \cite{bates_fitting_2015, pinheiro2017package, wilkinson1973symbolic} as:

\begin{minipage}[h]{\columnwidth}
{\hspace{-12pt}\includegraphics{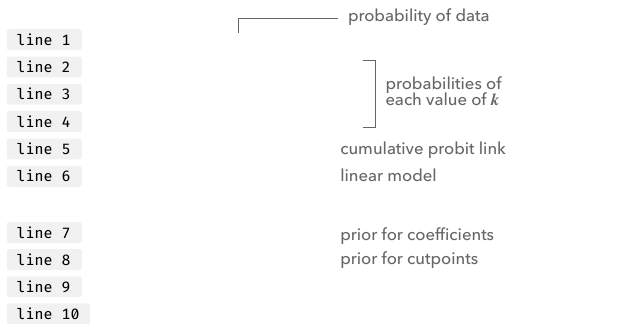}\vspace{-156.5pt}}
\begin{align*}
    & \hskip4em \mathit{y} \sim \text{Categorical}(p_i) \\ 
    & \hskip4em p_1 = q_1 \\
    & \hskip4em p_j = q_j - q_{j - 1} \hskip1em \text{for} \hskip0.5em 1 < j < K \\
    & \hskip4em p_K = 1 - q_{K - 1} \\
    & \hskip4em \mathrm{probit}(q_j) = \tau_j - \phi_i
    \\
    & \hskip4em \phi_{i} \hskip0.5em = \hskip0.5em \beta_{\textsc{a}[i]} + \beta_{\textsc{c}[i]} + \beta_{\textsc{i}[i]} + \\
    & \hskip7em \beta_{\textsc{a}[i]} \cdot \beta_{\textsc{i}[i]} + \beta_{\textsc{c}[i]} \cdot \beta_{\textsc{i}[i]} \\
    & \hskip4em \beta \hskip0.5em \sim \hskip0.5em \text{Normal}(0, 1) \\
    & \hskip4em \tau_j \hskip0.5em \sim \hskip0.5em \text{Normal}(0, 1) \\
    & \hskip4em i \in \{1...N\} \hskip4em (N \; participants) \\
    & \hskip4em k \in \{1...K\} \hskip3.8em (K \; \text{number of ordered categories}) \\[-8pt]
\end{align*}
\end{minipage}

\customsubsubsection{Line 1:} The rating the participants assigned is modeled as a Ordered categorical distribution. 

\customsubsubsection{Line 2-4:} The categorical distribution takes a vector of probabilities $p = {p_1, p_2, p_3, p_4, p_5, p_6}$ of probabilities of each response value below the maximum response ($p_7$). 

\customsubsubsection{Line 5:} Each response value k in this vector is defined by its link, using the probit function to an intercept parameter, $\tau_k$ and the linear model $\phi_i$.

\customsubsubsection{Line 6:} For each response, we assume that participants' ratings depend on the condition that the principle (or the combination of principles) the scenario is based on.

\customsubsubsection{Priors:} We use weakly-regularizing priors centered on zero (no effect) while permitting the possibility of significant distortion: $\beta, \tau_j \sim \mathrm{Normal}(0, 1)$.

\customsubsubsection{Implementation:} We implemented these models in \texttt{R 4.4.0}~\cite{RLang_2024} and \texttt{CmdStanR 0.8.0}~\cite{cmdstanr}. The model ran for four chains with 5,000 warmup samples and 5,000 post-warmup samples each, thinned by 4 for a final total sample size of 5,000. We assessed convergence using the Gelman-Rubin diagnostic ($\hat{R}$ = 1.00 for all population-level parameters, correlations and standard deviations) and the (bulk and tail) effective sample sizes ($\text{ESS}_{min} \approx 2,000$).

\end{document}